 \documentstyle[12pt,epsfig]{article}
\def\beq{\begin{equation}}   \def\eeq{\end{equation}}
\def\bea{\begin{eqnarray}}   \def\eea{\end{eqnarray}}

\newcommand{\gsim}{\lower.7ex\hbox{$
\;\stackrel{\textstyle>}{\sim}\;$}}
\newcommand{\lsim}{\lower.7ex\hbox{$
\;\stackrel{\textstyle<}{\sim}\;$}}

\newcommand{\GeV}{\,\mbox{GeV}}
\newcommand{\MeV}{\,\mbox{MeV}}

\newcommand{\bibit}[1]{\bibitem{#1}}

\begin{document}

\def\lsim{\mathrel{\rlap{\lower3pt\hbox{\hskip0pt$\sim$}}
    \raise1pt\hbox{$<$}}}         
\def\gsim{\mathrel{\rlap{\lower4pt\hbox{\hskip1pt$\sim$}}
    \raise1pt\hbox{$>$}}}         

\begin{titlepage}
\renewcommand{\thefootnote}{\fnsymbol{footnote}}

\begin{flushright}
JLAB-THY-99-13\\
UND-HEP-99-BIG\hspace*{.2em}06\\
hep-ph/9905520\\
\end{flushright}
\vspace{1.1cm}

\begin{center} \Large
{\bf
Theoretical Uncertainties in $\bf \Gamma_{\rm \!sl}( b \to u)$ 
}
\end{center}
\vspace*{.8cm}
\begin{center}
{\Large
Nikolai Uraltsev 
\\
\vspace{.5cm}
{\normalsize
{\it Jefferson Lab, 12000 Jefferson Avenue, Newport News, VA
23606,}\\
{\it Dept.\ of Physics,
Univ.\ of Notre Dame du
Lac, Notre Dame, IN 46556,}\\
{\it Petersburg Nuclear Physics Institute,
Gatchina, St.\,Petersburg, 188350 Russia
}
}
}
\vspace*{2.5cm}

{\footnotesize Contribution to Workshop on the Derivation of $|V_{cb}|$
and $|V_{ub}|$:\\ Experimental Status and Theory Uncertainties\\
{\it CERN, May 28 -- June 2, 1999}}

\vspace*{2.2cm}
{\large \bf Abstract}
\vspace*{.25cm}

\end{center}

\thispagestyle{empty}
\setcounter{page}{0}

\noindent
I review the existing theoretical uncertainties in relating the
semileptonic decay width in $b\to u$ transitions to the underlying
Kobayashi-Maskawa mixing element $|V_{ub}|$. The theoretical error bars
are only a few per cent in $|V_{ub}|$, with uncertainties
from the impact of the nonperturbative effects nearly negligible.

\end{titlepage}
\setcounter{footnote}{1}

\newpage

Inclusive semileptonic decay widths of beauty hadrons offer the
theoretically most clean way to determine the underlying KM mixing
angles describing the weak couplings of $b$ quark to $W$ boson. For
$b\to c$ transitions the semileptonic width is (almost) directly measured in
experiment. This allowed  one to determine $|V_{cb}|$ in a
model-independent way with unprecedented accuracy of only a few per
cent \cite{rev}.

Following the same route in the quest for $|V_{ub}|$ is much more
involved experimentally. Direct accurate measurement of 
$\Gamma_{\rm sl}(b\to u)$ for a long time seemed questionable. The
dedicated studies conducted over the last few years suggested a
feasibility of such measurements at the competitive level of
model-independence. The first results have already been reported
\cite{aleph}. Discussion of some theoretical aspects involved in 
unfolding the $b\to u$ width from the measurable decay 
distributions, can be found in Refs.\,\cite{bigdu}.

In this note I review the existing theoretical uncertainties in
relating $\Gamma_{\rm sl}(b\to u)$ to  $|V_{ub}|^2$. They are rather
small, and will be dominated by the uncertainties involved in the 
experimental determination of $\Gamma_{\rm sl}(b\to u)$ in the foreseeable
future.

\section{The theoretical framework}

The operator product expansion (OPE) applied to the inclusive decay
probabilities of a heavy hadron yields the basic expression 
\cite{buvprl} 
\beq
\Gamma_{\rm sl}^{b\to u}(B)\;=\; \frac{G_F^2 m_b^5}{192\pi^3}\,
|V_{ub}|^2 \left\{
A_0\left(1-\frac{\mu_\pi^2-\mu_G^2}{2m_b^2}\right) - 
2\frac{\mu_G^2}{m_b^2} + {\cal O}\left(\frac{1}{m_b^3}\right)
\right\}\;,
\label{1}
\eeq
with the leading power corrections given in terms of the two expectation
values
\beq
\mu_\pi^2\;=\; \frac{1}{2M_B} \langle B | \bar{b} (i\vec{D})^2b |B\rangle 
\;, 
\qquad \;
\mu_G^2\;=\; \frac{1}{2M_B} 
\langle B | \bar{b} {\small \frac{i}{2}}
g_s\sigma_{\mu\nu}G^{\mu\nu} b |B\rangle 
\label{3}
\eeq
having the transparent physical meaning. The practically most
important feature of the OPE result is the absence of the potential
$1/m_b$ nonperturbative corrections \cite{buvprl}. They could have been
naively expected from the strong dependence of $\Gamma_{\rm sl}(B)$ on the 
mass, and their potential size could have been as large as 20 to 40\%
even for $b$ particles. The actual nonperturbative corrections in QCD start
with $1/m_b^2$ and hence emerge at the scale of $5\%$. Therefore, in
practical terms they must be included, but can be treated in a
simplified fashion relying on the so-called `practical' version of the OPE
in QCD. The leading terms in Eq.\,(\ref{1}) representing the partonic
width free from bound-state nonperturbative effects, deserves attention to
the first place. It includes the purely perturbative corrections
embedded into the coefficient function $A_0$.

\subsection{Perturbative corrections}

The perturbative expansion of the decay width has the general form
\beq
\Gamma_{\rm sl}^{\rm pert}(b\to u) = \frac{G_F^2 m_b(\mu)^5}{192\pi^3}\,
|V_{ub}|^2 \left[
1 +  a_1\frac{\alpha_s(\tilde\mu)}{\pi} + 
a_2\left(\frac{\alpha_s(\tilde\mu)}{\pi}\right)^2 +
a_3\left(\frac{\alpha_s(\tilde\mu)}{\pi}\right)^3\,+...
\right] .
\label{5}
\eeq
The important feature explicit in Eq.\,(\ref{5}) is that in the quantum
field theory like QCD the masses, similar to all other `couplings'
determining the underlying Lagrangian, are ``running'' (depend on
normalization point). As a result, the perturbative coefficients $a_1$,
$a_2$,\ ... depend, in general, on the normalization points $\mu$ and
$\tilde\mu$ for $m_b$ and $\alpha_s$, respectively. 

The perturbative coefficient $a_1$ is well known from the computations
of the QED corrections to the muon decay. The effects associated with
running of $\alpha_s$ in the first-order QCD corrections generate the
whole series in $\alpha_s$ and they were computed in Ref.\,\cite{bbbsl} to all
orders in the generalized BLM \cite{blm} approximation. The most
challenging used to be the so-called ``genuine'' non-BLM $\alpha_s^2$
perturbative corrections, which were recently computed for $b\!\to\!u$ in 
Ref.\,\cite{timo}. The normalization scheme for the heavy quark mass
appropriate for the heavy quark expansion was found in 
Refs.\,\cite{five,blmope}. The two-loop (and the all-order BLM)
evolution of such mass was computed in  Ref.\,\cite{dipole}. Using this
low-scale running quark mass was instrumental for the precise numerical
evaluation of $m_b$,  Ref.\,\cite{my}. 

It is important to note that it is not possible to choose $\mu\!=\!0$ (which
would correspond to using the `pole' mass $m_b^{\rm pole}$ in
Eqs.\,(\ref{1},\ref{5})) to evaluate the width with a power-like
accuracy \cite{pole}: this would bring in uncontrollable $1/m_b$
corrections from the infrared domain of momenta $\sim \Lambda_{\rm
QCD}$. As mentioned above, such uncertainties would be at the level of
$20\%$ for $B$ decays.

The size of the higher-order perturbative corrections in  Eq.\,(\ref{5})
is crucial for evaluating the accuracy of our estimates of the
semileptonic width. The well-known source of the potentially large
higher-order perturbative effects is associated with the running of
$\alpha_s$. Since the leading BLM series has been computed, these 
effects cannot induce sizeable uncertainties. In the case of
semileptonic widths there is a potentially more significant source
associated with the large power $n\! =\! 5$ of $m_b$ in $\Gamma_{\rm sl}$.
This dependence can generally lead to the coefficients $a_k$ in the 
perturbative order $k$ growing as
$n^k$ \cite{five}. This is true even if
the strong coupling does not run at all. However, the leading subseries
of such corrections can be resummed. It reduces to choosing the proper
normalization point $\mu$ for the mass $m_b(\mu)$ in  Eq.\,(\ref{5}),
which must scale according to $\mu\sim m_b/n$ \cite{five}. 
Below I illustrate the utility
of this large-$n$ resummation on the example of the second-order
corrections. This complements the analysis of Ref.\cite{upset} dedicated 
to the BLM corrections.

Since the BLM subseries is known, it is convenient to concentrate on the
non-BLM corrections alone which are obtained subtracting the terms of the form
$\beta_0^k \left(\frac{\alpha_s}{\pi}\right)^{k+1}\!\!$, where
$\beta_0=\frac{11}{3} N_c-\frac{2}{3} n_f$ is the first coefficient of
the $\beta$-function for the strong coupling. ($n_f$ denotes the 
number of light flavors excluding $b$ quark). This is particularly
useful in discussing the second-order corrections since the
non-BLM coefficient $a_2^{(0)}$ does not depend on the scale $\tilde\mu$
chosen for the strong coupling $\alpha_s(\tilde\mu)$. 

In terms of the pole mass (that is, with $\mu \to 0$) the two non-BLM
coefficients $a_1^{(0)}$ and $a_2^{(0)}$ are \cite{timo}
\beq
a_1^{(0)}(0)\,=\, -\frac{2\pi^2}{3}+\frac{25}{6}\,\simeq\, -2.41\;, \qquad
a_2^{(0)}(0)\,\simeq\, 5.54\;.
\label{11}
\eeq
The literal value of $a_2^{(0)}(0)$ is significant. The analysis of
Ref.\,\cite{five} suggests, however, that the appropriate value of the
normalization point $\mu$ for the mass $m_b(\mu)$ is $\mu\approx m_b/n =0.2
m_b$. In other words, the perturbative coefficients $a_1^{(0)}(m_b/5)$, 
$a_2^{(0)}(m_b/5)$,\ ... must not be enhanced. For illustration
we can neglect them altogether, that is, assume they vanish.  
This would determine the perturbative 
coefficients in terms of the pole mass via the
relation \cite{dipole}
\beq
m_b(0) \simeq m_b(\mu) 
\!+\!
\left(\frac{16}{9} \mu \!+\!\frac{2}{3} \frac{\mu^2}{m_b}\right)
\!\left(\frac{\alpha_s}{\pi}
\!-\! N_c\left(\frac{\pi^2}{6}\!-\!\frac{13}{12}\right)
\left(\frac{\alpha_s}{\pi}\right)^2
\right)+ 
{\cal O} \! \left(
\left[\frac{\alpha_s}{\pi}\right]^3\!\!,\, \beta_0^k
\left[\frac{\alpha_s}{\pi}\right]^{k+1}
\right) .
\label{13}
\eeq
In this way with $\mu=0.2 m_b$ we arrive at the estimate 
\beq
a_1^{(0)}(0)\,\simeq \, -1.91 \;,
\qquad
a_2^{(0)}(0)\,\simeq\, 4.68\;.
\label{15}
\eeq
The magnitude of the second-order coefficient is reproduced! 

Clearly, an accurate match obtained above is partially accidental, and
{\it a priori} one should not have expected such an approximation to
work with better than $30\%$ accuracy. It is not possible to
rule out using, say, $\mu=0.2 m_b$ or $\mu=0.3 m_b$. In any case, even
complete  account for the $n$-enhanced corrections leaves out regular
terms, for example, a finite short-distance renormalization of the
$\bar{b}\gamma_\alpha(1\!-\!\gamma_5) u$ weak current characterized by the
momentum scale $\sim m_b$. These are not peculiar to the decay widths
and are expected to have perturbative expansions with coefficients of
order unity.

Thus, the calculated second-order perturbative corrections are very
moderate for $m_b$ normalized at the scale about $m_b/5\simeq 1 \GeV$,
and the similar behavior is expected from the higher-order effects
(first of all, $\alpha_s^3$ corrections). On the other hand, just this
low-scale running mass $m_b(\mu)$ can be accurately determined from the
$e^+e^- \to b\bar{b}$ cross section near the threshold \cite{mbee}, with
the most stability at $\mu\simeq 1.2\GeV$ (for a qualitative discussion
see, {\it e.g.}, Ref.\,\cite{varenna}, Sect.\,3.2). 

The value of the low-scale running mass $m_b(\mu)$ is routinely
translated to the scale $\mu=1\GeV$. Therefore, we express the width in
terms of the $b$ quark mass normalized at this scale:
$$
\Gamma_{\rm sl}^{\rm pert}(b\to u) \simeq  
\frac{G_F^2 m_b(1\GeV)^5}{192\pi^3}\,
|V_{ub}|^2 \left[
1 +  a_1(1\GeV)\frac{\alpha_s(\tilde\mu)}{\pi} + 
a_2^{(0)}(1\GeV)\left(\frac{\alpha_s(\tilde\mu)}{\pi}\right)^2 +
\right.
$$
\beq
\left.
\qquad\qquad \qquad \qquad\qquad \qquad 
+\;
a_2^{_{^{\rm BLM}}} (1\GeV, \tilde\mu)
\left(\frac{\alpha_s(\tilde\mu)}{\pi}\right)^2
\right]\, ,
\label{19}
\eeq
$$
a_1(1\GeV)\simeq -0.32\,, \qquad a_2^{(0)}(1\GeV) \simeq -1.28 \;.
$$
Here $m_b\simeq 4.6\GeV$ is assumed.

The value of $a_2^{_{^{\rm BLM}}}$ depends on the 
scale $\tilde\mu$. If $\tilde\mu =m_b$ then  
$a_2^{_{^{\rm BLM}}}\approx 3.6$ 
(neglecting the contribution from the $c\bar{c}$ 
quarks). 
Identifying the appropriate scale $\tilde \mu$ for $\alpha_s$ in the
standard $\overline{\rm MS}$ scheme, it is often advantageous to use its
commensurate scale for a more physical coupling, which to our accuracy 
amounts to using 
$\tilde \mu_{_{^{_{\,\overline{\rm MS}}}}} = {\rm e}^{-5/6} m_b$.
With this choice  $a_2^{_{^{\rm  BLM}}}
(1\GeV, {\rm e}^{-5/6} m_b) \simeq 4.8$. 
In view of a smallish size of the first-order correction in Eq.\,(\ref{19}) 
this has no practical significance, however.
The $c$ quark loops are expected to 
increase the value of $a_2^{_{^{\rm  BLM}}}$ by $1\pm 0.5$.

\subsection{Numerical value of $m_b(1\GeV)$}

The value of $m_b(1\GeV)$ can be accurately determined from
experimental cross 
section $e^+e^- \to b\bar{b}$. Even without incorporating any
corrections at all, one obtains a reasonable estimate about $4.60 \GeV$ 
(see, {\it e.g.}, \cite{varenna}).
The leading and next-to-leading perturbative corrections are modest and
yield an estimate $m_b(1\GeV) \simeq 4.50\; \mbox{to}\;4.55 \GeV$, depending 
on the value of $\alpha_s$. The NNLO determination
was performed recently \cite{my} resulting in   
$m_b(1\GeV) \simeq 4.56 \GeV$, with the stated uncertainty of
about $50\MeV$. There are certain directions along which  this estimate
can be refined in the future. On the other hand, a number of considerations
suggest that this value is on the lower side and most probably
represents the lower bound for the possible values of $m_b$. To be
conservative, I shall assign the uncertainty $60\MeV$ to $m_b$:
\beq
m_b(1\GeV)\, =\, 4.58\pm 0.06 \GeV\;.
\label{23}
\eeq
This value refers to the specific physical scheme (the ``kinetic'' mass)
defined in Refs.\,\cite{five,dipole}.

\section{Nonperturbative effects}
\subsection{Power corrections}

The $1/m_b^2$ nonperturbative corrections decrease the width by
approximately $-4\%$. The chromomagnetic expectation value is
estimated through the hyperfine splitting
\beq
\mu_G^2 \;\simeq\; \frac{3}{4} \left( M_{B^*}^2-M_B^2\right)\:\simeq\:
0.4\GeV^2\;,
\label{29}
\eeq
with the conservative estimated accuracy $\pm 25\%$ reflecting the
intrinsic $1/m_b$ effects and the perturbative corrections in the
coefficient function related to the complete field-theoretic definition
of this operator. 

The kinetic expectation value $\mu_\pi^2(\mu)$ traditionally is evaluated at
the scale $\mu\simeq 0.7\GeV$ and is somewhat uncertain at present. The
inequality $\mu_\pi^2(\mu) > \mu_G^2(\mu)$ \cite{motion,ineq,vcbopt} 
essentially
limits the range of its possible values; the critical review can be
found in Ref.\,\cite{rev}. Since the dependence of 
$\Gamma_{\rm sl}^{b\to u}$ on $\mu_\pi^2$ is weak, we adopt here an
overly conservative range
\beq
\mu_\pi^2 \; = \; \left( 0.6\pm 0.2\right) \GeV^2\;.
\label{31}
\eeq

At order $1/m_b^3$ the nonperturbative effects can show up as the
$1/m_b$-suppressed pieces of the kinetic and chromomagnetic operators,
which are not expected to exceed the $25\%$ level of their leading-order
contributions accessed above. There are also two new operators, one of
them leading to the so-called Darwin term
\beq
\rho_D^3\,=\, -\frac{1}{2M_B} \langle B | 
\frac{g_s^2}{2} \bar{b} \gamma_\alpha t^a b \, 
\sum_q \bar{q} \gamma_\alpha t^a q
| B \rangle \;.
\label{33}
\eeq
The second operator is of a similar four-fermion form but includes only
the $u$ light  quark and has a different color and Lorentz structure:
\beq
\frac{1}{2M_B} \langle B | 
\bar{b} \gamma_\alpha (1\!-\!\gamma_5)u \:  
\bar{u} \gamma_\beta (1\!-\!\gamma_5)b |
B \rangle  (\delta_{\alpha\beta}- v_\alpha v_\beta)\;, \qquad v_\mu=
\frac{P^B_\mu}{M_B}\;.
\label{34}
\eeq
The later operator generates the `spectator-dependent' corrections
sensitive to the flavor of the light antiquark in $B$ meson. The
expectation value Eq.\,(\ref{34}) describes the effect of weak
annihilation (WA) and in general differentiates 
$\Gamma_{\rm sl}^{b\to u}(B^-)$ from $\Gamma_{\rm sl}^{b\to u}(B^0)$; the
Darwin operator is an isosinglet and affects the widths uniformly.
The above operators emerge at the momentum scale governed by
$m_b$ and must be evolved down to the hadronic scale. They 
mix under renormalization, and different color and Lorentz
structures appear.

With massless leptons the effect of WA vanishes in the factorization 
approximation, and is expected to be dominated by the
nonfactorizable piece \cite{WADs}. Therefore, it must be
suppressed. The coefficient of the Darwin operator was first evaluated 
in Refs.\,\cite{volshif,bds} (see also \cite{grekap}), and its
expectation value can be estimated by factorization \cite{motion}. 
The detailed discussion can be found in the
dedicated paper \cite{four}, with the final estimate
\beq
\frac{\delta\Gamma_{\rm sl}^{{\rm Darwin}}(b\to u)}{\Gamma_{\rm sl}(b\to u)} 
\;\approx \; -(1 \div 2) \% \;.
\label{35}
\eeq

To get an independent idea of the size of nonfactorizable contributions, we can
use data on $D$ decays. In the estimates, we employ the following ingredients
as input:

$\bullet$ $SU(3)$ symmetry

$\bullet$ WA effect in $\Gamma_{\! D_s}\!\!-\!\Gamma_{\! D^0}$

$\bullet$ Nonperturbative effects in $\Gamma_{\rm sl}(D)$.

More specifically, we attribute a significant part of the excess in
$\Gamma_{\rm sl}(D)$ compared to the OPE estimate, to the effect of the
nonfactorizable four-fermion expectation value.\footnote{The relation of this
assumption to the possible duality violation will be elucidated elsewhere.}
We then arrive at the following evaluation of the isosinglet and isotriplet
effects, respectively:
\bea
\frac{\delta_s \Gamma_{\rm sl} (b\to u)}{\Gamma_{\rm sl} (b\to u)}
& \equiv &
\frac{\delta \Gamma_{\rm sl}^{{\rm nf}} (b\to u)\vert_{_{B^-}}
+ \delta\Gamma_{\rm sl}^{{\rm nf}} (b\to u)\vert_{_{B^0}}}
{2\Gamma_{\rm sl} (b\to u)} \approx 0.04 \,
\frac{\delta \Gamma_{\rm sl}^{{\rm nf}} (D^0)}{\Gamma_{\rm sl} (D^0)}
\nonumber
\\
\frac{\delta_{_{^{\rm WA}}} \Gamma_{\rm sl} (b\to u)}{\Gamma_{\rm sl}(b\to u)}
& \equiv & 
\frac{\delta \Gamma_{\rm sl}^{{\rm nf}} (b\to u)\vert_{_{B^-}}
- \delta \Gamma_{\rm sl}^{{\rm nf}} (b\to u)\vert_{_{B^0}}}
{\Gamma_{\rm sl} (b\to u)} \approx 0.05\,
\frac{\Gamma_{D_s}^{^{_{\rm WA}}} - 
\Gamma_{D^0}^{^{_{\rm WA}}} }{\Gamma_{D^0}}\;. \qquad
\label{36}
\eea
Here we have used the computations of Ref.\,\cite{four} and its 
estimates of the
so-called color-straight expectation values. 
The effects safely below a per cent level
are discarded.

According to the analysis of Ref.\,\cite{inst}, the missing fraction of the $D$
semileptonic width constitutes about $50\%$. (This corresponds to a small 
nonfactorizable piece in the expectation value, $g_s \sim 0.02$ \cite{WADs}.)
Keeping in mind that $m_c$  is not much larger than the
hadronic scale, it is reasonable to adopt
$$
\frac{\delta \Gamma_{\rm sl} ^{{\rm nf}} (D^0)}{\Gamma_{\rm sl} (D^0)}
\;\approx \;0.25 \mbox{ to } 0.5\;.
$$

The analysis of the second Ref.\,\cite{WADs} suggests that the major origin of
difference between $\tau_{_{\! D_s}}$ and 
$\tau_{_{\!D^0}}$ comes from WA, while 
other effects, {\it e.g.} related to  $SU(3)$ breaking probably do not exceed
$\sim \!\!5\%$. Therefore, using $\tau_{_{\!D_s}} / \tau_{_{\!D^0}} 
\simeq 1.20$ \cite{tauds} 
we assess
\bea
\frac{\delta_s \Gamma_{\rm sl} (b\to u)}{\Gamma_{\rm sl} (b\to u)}
&\approx &
(1 \div 2)\%
\nonumber
\\
\frac{\delta_{_{^{\rm WA}}} \Gamma_{\rm sl} (b\to u)}{\Gamma_{\rm sl}(b\to u)}
& \approx  & -1\%\;.
\label{37}
\eea
The isosinglet enhancement of the width tends to offset the effect of the
Darwin operator. 

The literal application of the $1/m_Q$ expansion in decays of charmed mesons is
questionable and is at best semiquantitative. Therefore, we view the above
computation rather as an evaluation of the significance of the potential
contributions. We will also allow for 
a factor of $2$ increase in the effects, to have a
more confident assessment of the related uncertainties. 
Let us recall that studying the $b\!\to \!u$ decay distributions for charged
and neutral $B$ {\it separately} will provide an important information
on the nonfactorizable effects in heavy mesons \cite{WADs}.

Thus, the nonperturbative effects in $\Gamma_{\rm sl}(b\!\!\to \!\!u)(B)$ 
computed in the OPE are
expected to be about $-5\%$, and can be reliably estimated.

\subsection{Violations of local duality}

The predictions based on the practical applications of the OPE applied
to most observables in the Minkowski space, first of all decay
rates, to a certain extent rely on local quark-hadron duality. Although
the conceptual origin of its possible violation at finite energies has
been clarified in recent studies \cite{shifcont,inst}, the reliable dynamic
evaluation of its significance at intermediate energies still lies
beyond the possibilities of modern theory. Can one expect duality
violations to affect credibility of determination of $|V_{ub}|$ from 
$\Gamma_{\rm sl}^{b\to u}(B)$?

For the OPE-amenable observables violation of local duality is
intrinsically related to the asymptotic nature of the power expansion in
QCD. This means that at finite $m_b$ including higher and higher terms
in $1/m_b$ -- even if they all were known -- would improve the accuracy
of the predictions only up to a point. This perspective may suggest {\it
a priori} an optimistic viewpoint that the duality violation is
safely below the effect of the nonperturbative corrections which have 
been evaluated. 
While this is the most natural assumption  which often holds, it must not
necessary be true, as can be traced in certain model considerations. 

Nevertheless, there are sound reasons to believe that local duality
violation in $\Gamma_{\rm sl}^{b\to u}(B)$ should not be noticeable at a
per cent level relevant in practice. Some of the arguments can be found
in the lectures \cite{varenna}, Sect.\,3.5.3\ and rely on the general
constraints the duality-violating effects must obey, on the one hand, and
on the experimental information on other hard processes in QCD at intermediate
energies, most notably the resonance physics. The key fact is that the
energy release is large enough, so that a significant number of channels are
open even if the resonance structures are not yet completely washed out
in a particular channel \cite{inst}.
The dynamical models of duality violations typically predict negligible
effects at the mass scale around $5\GeV$.

The violations of local duality in the decay widths was recently
considered in the framework of the exactly solvable 't~Hooft model --
$1\!+\!1$ dimensional QCD in the limit of a large number of colors. This model
exhibits in full the part of the actual QCD phenomenology which is
expected to play a crucial role in violation of local duality, {\it
viz.}\ quark confinement and manifest resonance dominance. The
analytic studies in Refs.\,\cite{d2,d2wa} led to the conclusion that for
the actual mass of the $b$ quark the duality-violating effects must lie
below a percent level. The similar conclusion can be drawn from the
numerical studies of  Refs.\,\cite{gl} viewed from the proper
perspective.

Of course, QCD in $1\!+\!1$ dimensions cannot fully represent ordinary QCD.
In particular, it misses to incorporate the possible effects of
transverse gluons absent in two dimensions. In this respect, it is
important to keep in mind that the overall effect of the perturbative
corrections in  $\Gamma_{\rm sl}^{b\to u}$ in the proper OPE approach
does not exceed a $10\%$ level, see Eq.\,(\ref{19}), so that even a
delayed onset of duality here can hardly bring in a significant effect.

To summarize, one expects duality violation to be negligible in
$\Gamma_{\rm sl}(b\to u)$.

\subsection{Comments on the literature}

Since the development of the dynamic $1/m_Q$ expansion for the
inclusive widths, suggestions surface every now and then in the
literature which challenge applicability of the OPE in one form or
another. For example, paper \cite{jin1} claimed identifying a certain
class of `kinematic' nonperturbative effects which are allegedly missed
in the conventional OPE approach, and must be incorporated additionally.
This was applied to the semileptonic $b\to u$ width in Ref.\,\cite{jin2}
and resulted in an essentially larger positive nonperturbative
corrections. It thus seems appropriate to dwell on this issue in the
context of the present note.

Unfortunately, there is certain misleading element in that the approach of
Refs.\,\cite{jin1,jin2} is claimed to be derived from the first
principles of QCD. This is not so in reality, and simply cannot be
since the result is incompatible with a few very general properties. In
particular, this refers to Eq.\,(2) of Ref.\,\cite{jin2} which is the
starting expression for the width. In fact, this is simply a parton model
motivated ansatz employing only the heavy quark analogue of the leading-twist 
distribution function. While the leading-twist distribution function
captures properly the major effects of the so-called ``Fermi motion'' 
\cite{alialt} on
the decay {\it distributions}, it cannot be responsible for the
corrections to the integrated rates which start explicitly only with 
the higher-twist effects. 

The model ansatz of Ref.\,\cite{jin2} is adjusted to correctly reproduce
the absence of $1/m_b$ corrections to the inclusive width. Its
deficiency nevertheless manifests itself already at the order $1/m_b^2$
where the first nonperturbative corrections to the width appear. Indeed,
it is easy to obtain what is the $1/m_b$ expansion of the ansatz, 
using Eqs.\,(5) and (6) of Ref.\,\cite{jin2}:
\beq
\Gamma_{\rm sl}^{ {\rm Ref.[31]}}(B)\;=\;
\Gamma_{\rm sl}^{\rm parton} \, \left[
1+\frac{35}{6} \frac{\mu_\pi^2}{m_b^2} -\frac{5}{2}
\frac{\mu_G^2}{m_b^2} + {\cal O}\left(\frac{1}{m_b^3} \right)
\right]\;.
\label{51}
\eeq
The comparison with the OPE result Eq.\,(\ref{1}) shows that neither
the chromomagnetic nor the kinetic operator contributions are
reproduced. The most dramatic difference appears in the latter: the
coefficient for $\mu_\pi^2$ is almost 12 times larger and has the
opposite sign!

It is a simple matter to see which of the two expressions is correct,
and this does not require going through the whole machinery of the OPE in QCD.
It suffices to look at the decay rate of a {\em free} quark moving with
the small velocity $\vec{v} \sim 1/m_b$. Its spacelike momentum is then 
$\vec{p}=m_b \vec{v}$, and the decay rate is simply suppressed by the
Lorentz dilation factor $(1+\vec{p}^{\:2}/m_b^2)^{-1/2} \simeq
1-\vec{p}^{\:2}/2m_b^2$. This is the meaning of the corresponding
OPE correction in Eq.\,(\ref{1}) \cite{WADs,motion}. Clearly,
Eq.\,(\ref{51}) stands no chance to hold for $B$ mesons if it fails
so heavily even for a free particle. 

As a matter of fact, there is a nontrivial example of a theory where
the semileptonic decay width of the strongly interacting confined $b$
quark can be calculated analytically and confronted to the $1/m_b$
expansion derived in the OPE \cite{d2}. This is the 't~Hooft model
mentioned in the previous section. The theoretical computations are
usually compounded by the necessity to evaluate in parallel the
perturbative corrections to the width. However, in the special case of
vanishing lepton masses, the perturbative corrections in the 't~Hooft
model can be computed to all orders in perturbation theory
\cite{d2}. In this case the decay width of the $B$ mesons, both in
$b\to u$ and $b\to c$ transitions, was shown to coincide with its OPE
expansion at all computed orders in $1/m_b$. Incidentally, the exact
expression for the width in terms of the corresponding light-cone
wavefunction of the $B$ meson would have an essentially 
{\it different} functional
form from the ansatz postulated in Ref.\,\cite{jin2}.

Failing to comply QCD already at the level of the leading
nonperturbative effects, the ansatz of Ref.\,\cite{jin2} intrinsically
contains large $1/m_b^3$ and higher-order terms, which likewise have
nothing to do with actual strong interactions. At the same time, it offers
no room for the actual $1/m_b^3$ effects, the potential
spectator-dependent contributions from WA which have the transparent
underlying origin.

Considering the above facts, we have to conclude that in what concerns
the integrated rates, the approach of Refs.\,\cite{jin1,jin2} is
fundamentally flawed, and its numerical outcome cannot be used even to
get model-dependent insights into 
possible theoretical uncertainties associated with 
accounting for the bound state and hadronization dynamics.

\section{Summary and conclusions}

Assembling all pieces together, we evaluate the present theoretical
predictions for $\Gamma_{\rm sl}^{b\to u}(B)$ as follows:
\beq
\Gamma_{\rm sl}^{b\to u}(B) = 66 {\rm ps}^{-1} |V_{ub}|^2 
\left[
1 \! +\! 0.065 \frac{m_b(1\GeV) \! -\! 4.58\GeV}{60\MeV}
\pm 0.02_{\rm pert} \pm 
0.035_{\rm nonpert}\,
\right] , 
\label{61}
\eeq
where the last term lumps together the uncertainties in accounting for 
the nonperturbative effects. The largest source of uncertainty remains
in the precise value of the running $b$ quark mass, although it is only
a few per cent. For $|V_{ub}|$ itself we then arrive at 
\beq
|V_{ub}|=0.00442 \left(\frac{{\rm BR}(B^0\rightarrow
X_u\ell\nu)}{0.002}
\right)^{\frac{1}{2}}\left(\frac{1.55\,\rm
ps}{\tau_B}\right)^{\frac{1}{2}}
\cdot \left(1 \pm 0.025_{_{\rm QCD}} \pm 0.035_{m_b}
\right)
\;.
\label{63}
\eeq

The uncertainty in $\mu_\pi^2$ does not affect the
theoretical predictions at an appreciable level. No significant
uncertainty is expected through the uncalculated higher-order
perturbative effects. 
Some variation, in principle, can be allowed for from the
precise value of the strong coupling $\alpha_s$ at a few $\GeV$ scale.
While the $Z$-peak physics seems to yield a larger value of $\alpha_s$,
certain low-energy phenomenology would favor a lower value of
$\Lambda_{\rm QCD}$. Therefore, it may be premature to rely on the
larger value of $\alpha_s$ often applied to low-scale physics, see 
Ref.\,\cite{dok}. The literal dependence on the value of $\alpha_s$ of the 
width in Eq.\,(\ref{19}) is very weak, so it cannot be used to 
estimate the overall uncertainties in the perturbative corrections.

The current estimated value of $|V_{ub}|$ is close to the original
evaluation made in Ref.\,\cite{upset}.\footnote{Some numerical difference 
is related to using there a lower value of the quark mass.} 
This is not accidental, for both
the progress in the determination of $m_b$ and the state of the art
computations of the second-order corrections yielded very moderate
effects if one relies on the proper OPE-compatible low scale mass $m_b$, 
as in Ref.\,\cite{upset}. Such a stability is a good sign
indicating that the theoretical relation between 
$\Gamma_{\rm sl}^{b\to u}(B)$ and $|V_{ub}|^2$ rests on sound grounds. 

It is often tempting to determine the ratio of the KM mixing angles
$|V_{ub}/V_{cb}|$. When extracted from the decay widths,
$|V_{ub}|$ and $|V_{cb}|$ often share common theoretical uncertainties
which can partially cancel in the ratio. For example, this happens with the
dependence on the exact value of $m_b$. Nevertheless, it is important to
keep in mind that even the underlying problems are not always identical.
The energy release in $b \!\to \! u$ is safely large to ensure a good control
of the nonperturbative effects, without a recourse to the heavy quark
symmetry. A more limited energy release in the decays $b \!\to \! c$ makes it
{\it a priori} more vulnerable to possible effects of duality violation,
sensitive to the structure of the higher-order 
nonperturbative corrections and
to applicability of the heavy quark symmetry to charm particles through
the value of $m_b\!\!-\!\!m_c$. On this route the dominant dependence on
$\mu_\pi^2$ emerges for $|V_{cb}|$. All these theoretical ingredients
will be critically examined when the new generation of experimental data
on $B$ decays become available.

\vspace*{.2cm}

{\bf Acknowledgments:} \hspace{.1em} 
I am grateful to I.\,Bigi for collaboration, to K.\,Melnikov 
for useful discussions and to M.\,Shifman, M.\,Voloshin and to many 
experimental colleagues  
for their encouraging interest.
This work was supported in part by the National Science Foundation under
the grant number PHY~96-05080 and by the RFFI under the grant number
99-02-18355.

\end{document}